\documentclass{elsart}
\usepackage{graphics}
\begin{document}
\begin{frontmatter}

\title{Magnetic extrema in electronic susceptibility and
heat capacity of mesoscopic systems}
\author[Radium]{N.K. Kuzmenko\corauthref{cor}},
\corauth[cor]{Corresponding author.}
\ead{Kuzmenko@NK9433.spb.edu}
\author[Univ]{V.M. Mikhajlov}
\address[Radium]{V.G.Khlopin Radium Institute, 194021
St.-Petersburg, Russia}
\address[Univ]{Institute of Physics St.--Petersburg State
University 198904, Russia}

\begin{abstract}
Oscillating behaviour of the susceptibility $\chi$ and
heat capacity $C$ is considered for normal
and superconducting mesoscopic systems (nanoclusters and quantum dots).
It is proved that at low temperature an increasing
magnetic field applied to a mesoscopic
system generates local extrema of $\chi$ and $C$.
A {\em maximum\/} for $\chi$ and a
{\em minimum\/} for $C$ simultaneously arise in
those points of the field where crossings of quantum levels of
the normal and superconducting mesoscopic systems take place.
\end{abstract}

\end{frontmatter}

\section{Introduction}
Low temperature oscillations of diverse electromagnetic and
thermal characteristics (resistivity, magnetization,
susceptibility, thermal conduction, heat capacity etc.) of
macroscopic and mesoscopic bodies in an increasing uniform
magnetic field are apparently a general property of systems formed
by charged particles. These oscillations have been last years
studied experimentally and theoretically in mesoscopic systems of
different geometry and microscopical structure: in mesoscopic
rings~\cite{moskaletz}, in a thin spherical layer~\cite{aristov},
in superconducting
discs~\cite{buisson90,deo97,geim98,deo2000,schechter}, in
micron-size heterostrucures~\cite{dikin}. Observing such variety
of properties and objects one is forced to arrive at a conclusion
that there must be a common reason for arising this phenomenon. In
this Letter as a source of such oscillations we regard crossings
of the quantum levels of a mesoscopic system abstracting from the
physical nature of the levels and system itself. We concentrate
here on static properties (the magnetic susceptibility $\chi$ and
heat capacity $C$) in the regime of the thermodynamical
equilibrium of the canonical ensemble at low temperatures.

We assume that a system exposed to a uniform magnetic field~($B$)
possesses a number of discrete quantum states altering their mutual
energy disposition depending on $B$. Thereby for each value of
$B$ there exist a lowest state, the ground
state for a given $B$, and the nearest to it first excited state
with energies $E_0(B)$ and $E_1(B)$ respectively.
Each of these energies is a smooth function of $B$, however
$E_0(B)$ and $E_1(B)$ can locally have opposite slopes with $B$.
Therefore the difference $E_{10}(B)$
\begin{equation}
E_{10}(B)=E_1(B) - E_0(B); \; \;  E_{10}(B_0)=0.
\end{equation}                                             
can become equal to zero for a value of $B=B_0$.

In Sec.2 we show that this point of
the level crossing is a peculiar point where at very small temperatures
both $\chi$ and $C$ reach local extrema: a {\em maximum\/}
for $\chi$ and a {\em minimum\/} for $C$.

\section{Extrema in $\chi$ and $C$}
In the low temperature limit only two first terms practically
exhaust the canonical partition function
\begin{equation}
Z=exp\lbrack -\beta E_0(B)\rbrack + exp\lbrack -\beta E_1(B)\rbrack
+ Z_{\mathrm{r}},\;\;\;\beta=1/T.
\end{equation}                                            
The rest, $Z_{\mathrm{r}}$, gives a weak
background for
$\chi$ and $C$ near the point $B_0$ in the vicinity of which the
partition function can be expressed via a small parameter $\xi$:
\begin{eqnarray}
\xi=\frac{1}{2}\lbrace exp\lbrack -\beta E_{10}(B)\rbrack -1\rbrace
; \;\; \xi(B_0)=0, \\
\ln Z=-\beta E_0(B) + \ln(1+\xi) + \alpha + \ln2, \nonumber \\
\alpha=\frac{1}{2}\sum_{i\geq 2}exp\lbrace -\beta
(E_i -E_0)\rbrace\ll 1. \nonumber
\end{eqnarray}                                 

The magnetic  susceptibility and heat capacity are naturally
divided near $B_0$ into smoothly varying parts
$\chi_{\mathrm{s}}$, $C_{\mathrm{s}}$
and the rapidly varying functions $\chi_{\mathrm{e}}$,
$C_{\mathrm{e}}$ with extrema in $B_0$:
\begin{eqnarray}
\chi (B_0)  =  \chi_{\mathrm{s}} + \chi_{\mathrm{e}}, \\
\chi_{\mathrm{s}}  =  -\frac{1}{V}\frac{\partial^2(E_0 -\alpha T)}
{\partial B_0^2},
\;\;\;
\chi_{\mathrm{e}}  =  \frac{1}{\beta V}\frac{\partial^2\ln(1+\xi)}
{\partial B_0^2},
\end{eqnarray}                                   

\begin{eqnarray}
~~C(B_0)  =  ~~C_{\mathrm{s}} +~~C_{\mathrm{e}},  \\
~~C_{\mathrm{s}}  =  \beta^2\frac{\partial^2\alpha}
{\partial\beta^2}, \;\;\;
~~C_{\mathrm{e}}  =  \beta^2\frac{\partial^2\ln(1+\xi)}
{\partial\beta^2}.
\end{eqnarray}                                   
$V$ is the volume of the system. The derivative
$\partial/\partial B_0$ implies hereafter the derivative with
respect to $B$ in the point of $B=B_0$.

At low temperatures ($T<\delta$, $\delta$ is the mean
level spacing at $B=0$) we take into account only those
terms in Eqs.(5) and (7) which are proportional to the maximum
power of $\beta$ at $B=B_0$,
$\partial^n\xi/\partial B_0^n\simeq 0.5(-\beta)^n (\partial
E_{10}/\partial B_0)^n$, that gives
\begin{eqnarray}
\chi_{\mathrm{e}}(B_0)= \frac{\beta}{4V}\left(\frac{\partial
E_{10}}{\partial B_0}\right)^2>0, \\
\frac{\partial\chi_{\mathrm{e}}}{\partial B_0}=0,\;\;\;
\frac{\partial^2\chi_{\mathrm{e}}}{\partial B_0^2}=
-\frac{\beta^3}{8V}
\left( \frac{\partial E_{10}}{\partial B_0} \right)^4<0,\\
C_{\mathrm{e}}(B_0)=0,\\
\frac{\partial C_{\mathrm{e}}}{\partial B_0} = 0,\;\;\;
\frac{\partial^2C_{\mathrm{e}}}{\partial B_0} = \beta^2 \left(
\frac{\partial E_{10}}{\partial B_0}\right)^2>0.
\end{eqnarray}                                           

$\chi_{\mathrm{e}}$ is paramagnetic ($\chi_{\mathrm{s}}$ determines
the diamagnetic component of $\chi$) and
proportional to the squared difference
of the magnetic moments of the crossing states. Thus,
Eqs.(8)~-(11) confirm the existence of a local maximum in $\chi$
and a minimum in $C$ at the level crossing point $B_0$.
Increasing the field can generate appearance of new level
crossing points that will give $\chi$ and $C$ an oscillating functions
of $B$. Intervals between peaks and their amplitudes depends on
the physical nature of the system and its shape. Here we limit
ourselves by considering weak fields (the cyclotron radius is much
larger than the size of the system). The alterations of quantum
levels v.s. $B$ are caused by removing the spin and rotational
degeneracy, the latter occurs when a system has rotational symmetry.
To investigate oscillating behavior of $\chi$ and $C$ in Sec.3
we apply the independent electron model to normal mesoscopic systems
and in Sec.4 superconducting clusters of
the perfect spherical shape are described by using the exact solution
of the conventional superconducting hamiltonian.

\section{Magnetic oscillations in normal clusters}

The character of oscillations in electron $\chi$ and $C$ of a
mesoscopic system is dependent on the its shape. In spherical
clusters at $B=0$ each many--fold degenerate level (the
spin-orbital interaction is omitted here) displays a sheaf-wise
splitting. Hence magnetic sublevels originated from one spherical
$l$-shell ($l$ is an orbital momentum) cannot cross each other
with the growth of $B$. Thus the minimum strength of a magnetic
field, $B_{\mathrm{min}}$, (here we consider the case of the
closed Fermi--shell at $B=0$) required to arouse the first
crossing and the first extremum is determined by
$\delta_{\mathrm{sh}}$, the distance between the Fermi--shell,
$l_F$--shell, and the unoccupied $l_{F+1}$--shell at $B=0$:
$\mu_B^*B_{\mathrm{min}}(l_F+l_{F+1}+2)=\delta_{\mathrm{sh}}$,
$\mu_B^*=\mu_Bm^*/m$, $m^*$ is the effective electron mass. We
assume that orbital and spin projections are aligned at the first
crossing. Nevertheless the absolute value of $B_{\mathrm{min}}$ is
practically independent of $l$, since the level splitting near $F$
at $B=0$ is proportional to $l$: $\delta_{\mathrm{sh}}\simeq
(2l_F+1)\delta$, where $\delta=4\varepsilon_F/3N$ is the mean
level spacing of the Fermi gas near the Fermi energy
($\varepsilon_F$). Thus, $B_{\mathrm{min}}\sim \delta m^*/\mu_Bm$
and for nanometer normal metallic clusters ($N\simeq 10^5$,
$\varepsilon_F\sim 10$ $eV$ and $m=m^*$) $B_{\mathrm{min}}$ is of
the order of $1T$. The evolution of some single--electron levels
in a spherical cavity and corresponding extrema in $\chi$ and $C$
are shown in Fig.1.
\begin{figure}[p]
\scalebox{0.5}{\includegraphics{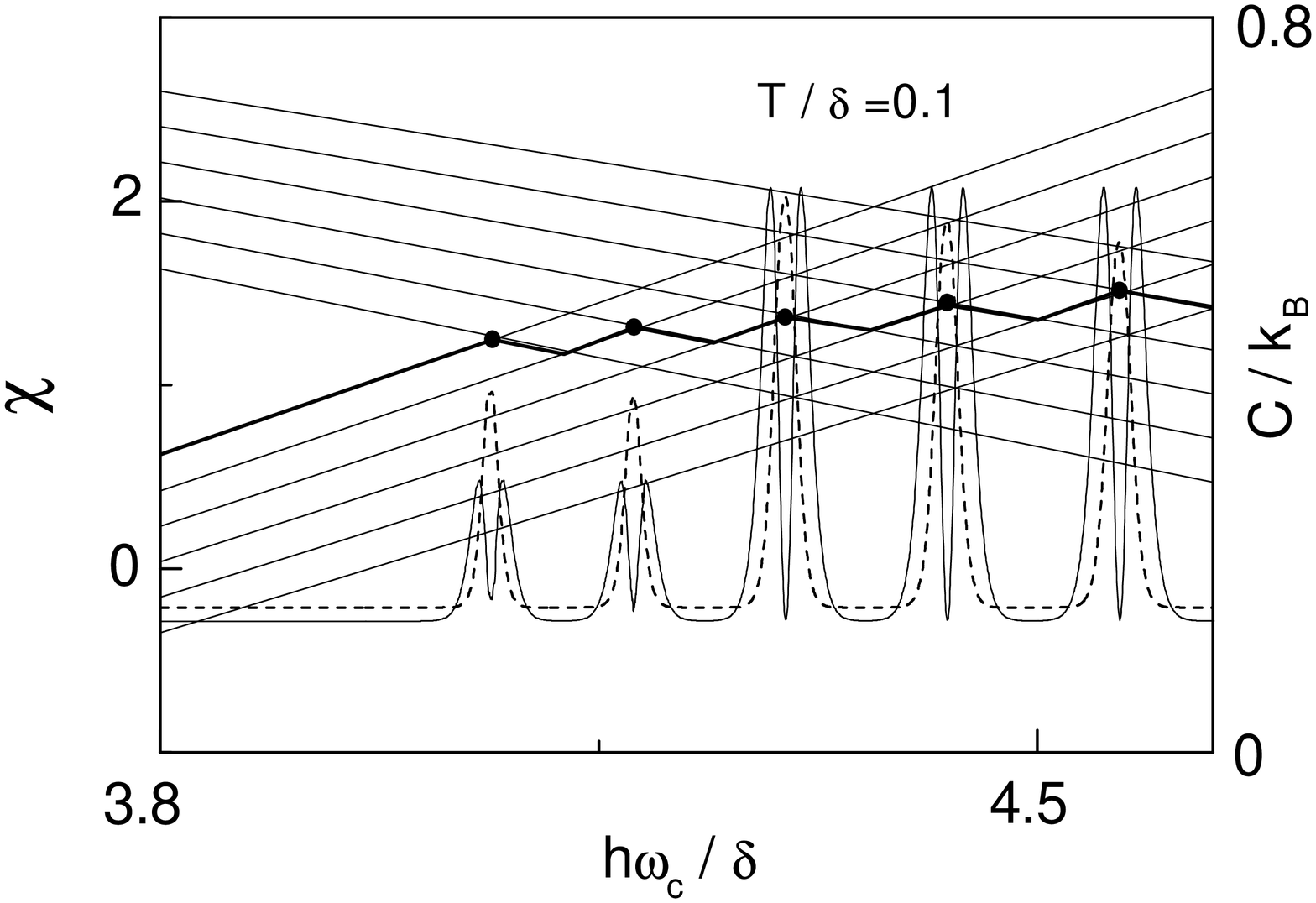}} \caption {Electron
levels vs the cyclotron frequency ($\hbar\omega_c=2\mu^*_BB$,
$\mu^*_B=\mu_Bm^*/m$) and magnetic oscillations of the
susceptibility $\chi$ (dotted line) and heat capacity $C$ (solid
line) for a sphere with~$N\simeq 10^5$. Points of level crossings
where $\chi$ and $C$ reach local extrema are marked by solid
circles. $\delta=4\varepsilon_F/3N$. The bold line corresponds to
the Fermi-level. $\chi$ is given in units
$10^4\cdot\mid\chi_{\mathrm{L}}\mid$, $\chi_L=-\mu_B^2
N/2V\varepsilon_F$ is the Landau diamagnetic susceptibility of the
degenerate free electron gas, $m=m^*$.}
\end{figure}

Properties of quantum dots are frequently interpreted in the
two-dimensional cylindrically symmetric oscillator
model~\cite{kouw}. This symmetry also gives a rather high
degeneracy of electron levels near the Fermi level, which is
removed in the same manner as in the previous spherical case.
$B_{\mathrm{min}}$ of the first level crossing is assessed
analogously: $B_{\mathrm{min}}\sim\varepsilon_F m^*/2N\mu_Bm$, $N$
is the free electron number in the quantum dot. As $\varepsilon_F$
and $m^*/m$ for such systems can be small (e.g. for GaAs
$m^*/m\sim 0.07$, $\varepsilon_F\sim 10$ $meV$ ~\cite{kouw})
oscillations of $\chi$ could be observable in relatively weak
fields for a wide range of $N$. $\chi$ and $C$ v.s. the cyclotron
frequency are shown in Fig.2,3 for $N\simeq 10^3$ ( we consider
the case of the closed cylindrical Fermi--shell at $T=B=0$ ). The
feature of $\chi$ and $C$ oscillations in systems with cylindrical
or spherical symmetry is the presence of several groups of peaks
corresponding firstly crossings of levels from two nearest
cylindrical or spherical shells separated at $B=0$ by one shell
gap, then cross levels separated at $B=0$ by two shell gaps and so
on.
\begin{figure}[p]
\scalebox{0.5}{\includegraphics{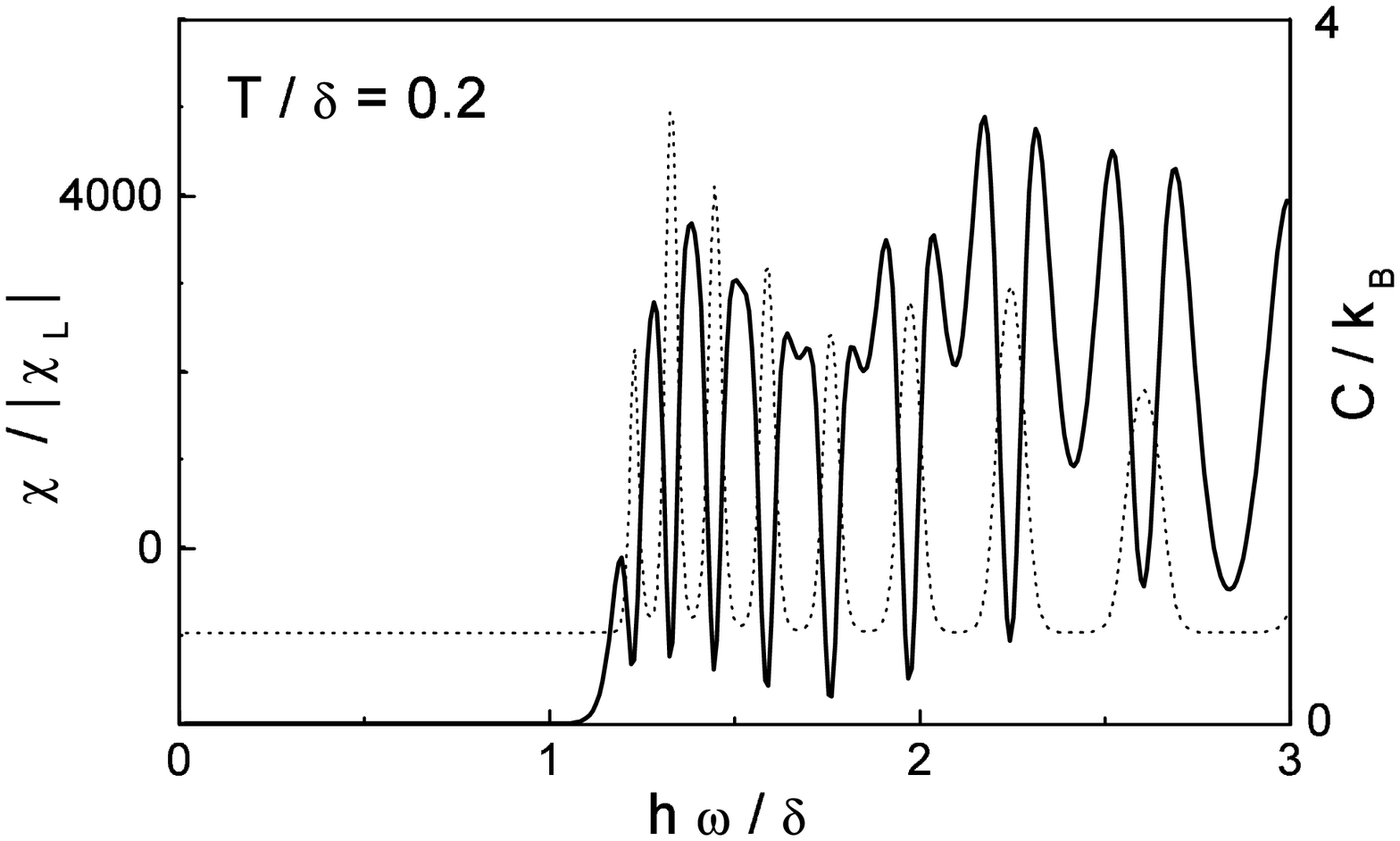}} \caption {Low
temperature magnetic oscillations $\chi$ (dotted line) and $C$
(bold line) in a two-dimensional cylindrical oscillator with
$N\simeq 10^3$. $\delta=4\varepsilon_F/3N$. $\chi$ is given in
$\chi_{\mathrm{L}}$.}
\end{figure}
Increasing temperature (Fig.3) leads to confluence of
separate peaks
so that curves for $\chi$ and $C$ become similar and
their gross structure is caused by the shell structure of
the single--particle energy spectrum.
\begin{figure}[p]
\scalebox{0.5}{\includegraphics{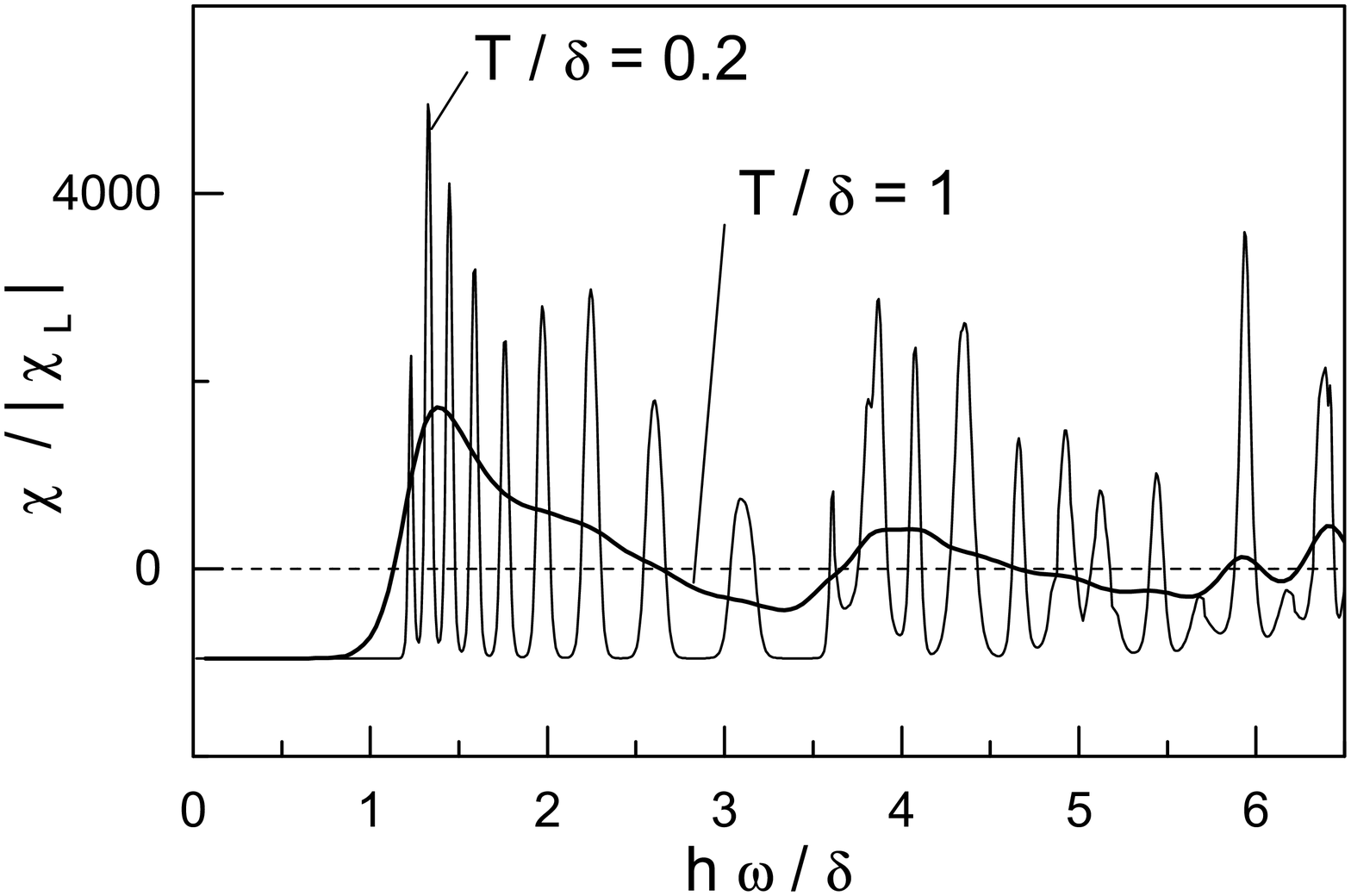}} \caption {$\chi$
vs the cyclotron frequency $\omega_c$ at different temperatures
for a two-dimensional cylindrical oscillator
 with~$N\simeq 10^3$ (closed Fermi-shell)
$\delta=4\varepsilon_F/3N$.}
\end{figure}

Much simpler behavior of $\chi$ and $C$ as function of $B$ is
observed in clusters having no rotation symmetry axis. In this
case level crossings are caused by the Zeeman splitting and
perturbation of orbital motion by weak fields proves to be
insignificant, because orbital energy shifts for such systems are
quadratic in $B$. Then signs of energy denominators appearing in
perturbative calculations are opposite for levels near the
Fermi--level that reduces second order terms. The example of such
``spin''  oscillations of $\chi$ is concerned with electron system
in the oscillator well with oscillator frequencies
$\Omega_x=0.9\Omega_y=0.5\Omega_z$ that corresponds to an
ellipsoid with semiaxes $a_x\simeq 1.1a_y=2a_z$ (we adopt the
relation between semiaxes and oscillator frequencies from
Ref.~\cite{ring}: $\Omega_xa_x=\Omega_ya_y=\Omega_za_z$). Such
system has no rotation axis and a magnetic field directed along
$x$--axis generates ``spin'' oscillations (Fig.4).
\begin{figure}[p]
\scalebox{0.5}{\includegraphics{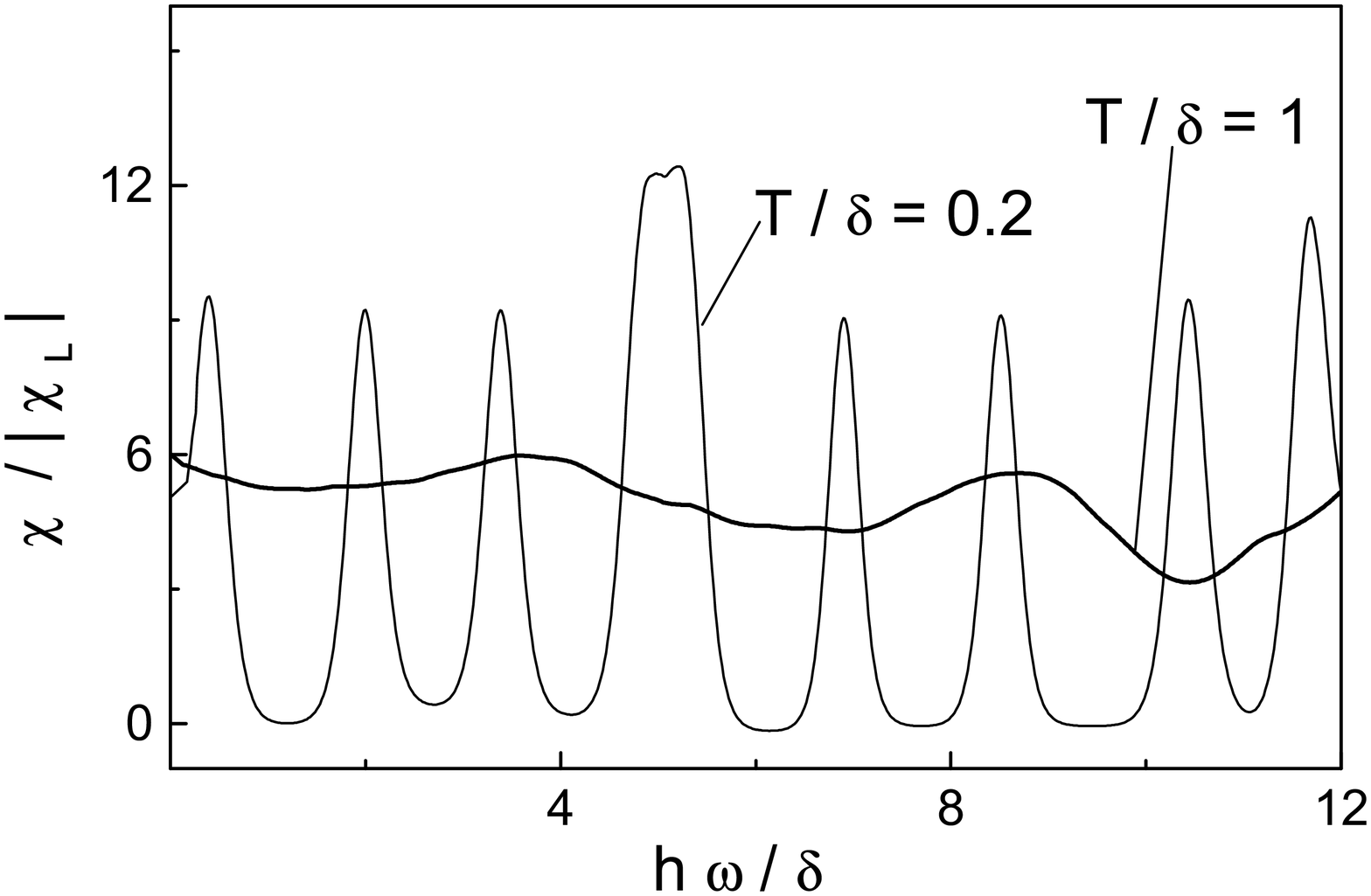}} \caption {$\chi$
vs $\omega_c$ at different temperatures for an anysotropic
oscillator ($\Omega_x=0.9\Omega_y=0.5\Omega_z$) with $N\simeq
10^5$, magnetic field is directed along $x$ -axis.}
\end{figure}
These
oscillations of $\chi$ in the anysotropic oscillator are much like
oscillations in the equal level spacing model \cite{denton}.

\section{Magnetic oscillations in the gapless
superconducting region} Another example of magnetic oscillations
of $\chi$ and $C$ is given by the single-shell model (SSM) of
superconductivity. Such model can be a reasonable approach to a
description of spherical superconducting clusters with numbers of
conduction electrons $10^3<N<10^5$ \cite{kuzmenko}. There is a
wide range of orbital momentum $l_F$ for states successively
becoming the Fermi level with growth of $N$. Averaging through $N$
gives $\widetilde l_{\mathrm{sh}}\simeq N^{1/3}$ \cite{kuzmenko}.
At $B=0$ the level spacing between degenerate shells
$\delta_{\mathrm{sh}}\simeq 2\widetilde l_F\delta\simeq
8\varepsilon_F/3N^{2/3}$ is much more than the BCS superconducting
gap at $B=T=0$, $\Delta_{\mathrm{BCS}}(0)$, e.g.
$\delta_{\mathrm{sh}}>10$~$meV$ and $\Delta_{\mathrm{BCS}}(0)< 1
$~$meV$ for $Al$-clusters with $N$ under consideration. Therefore
the superconducting pair correlations can arise between electrons
(their number is $N_{\mathrm{sh}}$) occupying only one shell with
a large $l$.

The Hamiltonian of SSM consists of the pairing interaction with
the strength $G$ and the interaction with a field $B$. The exact
spectrum of such Hamiltonian is well known~\cite{ring} and
at $B=0$ is determined by the seniority $\nu$
(unpaired electron number)
\begin{equation}
E_{\nu}=-\frac{G}{4}(N_{\mathrm{sh}}-\nu)(4l+4-N_{\mathrm{sh}}-\nu).
\end{equation}                                               
For each $\nu$ there is a multiplet of states (degenerate at
$B=0$) with different orbital and spin projections which are
splitted by a field. The energy of the lowest state for each
$\nu$ is
\begin{equation}
E_{\nu}(\omega)=E_{\nu} -
\frac{\omega}{4} \{ \nu(4l+2-\nu)-
\frac{1}{2}[1-(-)^{\nu}] + 4(1-\delta_{\nu.0} ) \},
\end{equation}                                                
Thus with growth of $B$ (in this Section
$\omega\equiv\hbar\omega\mu_B^*B$) the role of the
ground state is successively played by states with
$\nu=0,2,4,\ldots$ for even $N_{sh}$ or $\nu=1,3,5\ldots$ for odd
$N_{\mathrm{sh}}$, i.e. increasing $B$ leads to a sequence of
crossings of the $\nu$-level with $\nu+2$-level, resulting in
a step--wise
decrease of the pairing gap (the region of the gapless
superconductivity).

To take into account the temperature dependence of $\chi$ and $C$
quantities we have analytically summed up those components of the
canonical partition function $Z$ which involve states inside
the considered $l$--shell
\begin{eqnarray}
Z=\sum_{\nu} exp(-\beta E_{\nu})
[\Phi_{\nu}-\Phi_{\nu-2}(1-\delta_{\nu .0})],\\
\Phi_{\nu}=\sum_{\sigma_{\mathrm{min}}}^{\nu /2} \frac{2}{
1+\delta_{\sigma .0} }
\widetilde\Phi_{\nu/2+\sigma}\widetilde\Phi_{\nu/2-\sigma}
\cosh(2\beta\omega\sigma), \nonumber \\
\sigma_{min}=\frac{1}{4}[1-(-)^{\nu}], \nonumber \\
\widetilde\Phi_k=\delta_{k.0}+(1-\delta_{k.0}) \prod_{\mu=1}^k
\frac{\sinh(\beta\omega\frac{2l+2-\mu}{2}) }
{\sinh(\beta\omega\frac{\mu}{2}) }, \;\;\;\;
\beta=\frac{1}{k_BT}. \nonumber
\end{eqnarray}                                                  

Fig.5 shows $\chi$ and $C$
as functions of the magnetic field and temperature.
The latter is measured in $T_{\mathrm{crit}}$,
the critical temperature of the conventional
BCS theory for half--filled shell.
\begin{figure}[p]
\scalebox{0.5}{\includegraphics{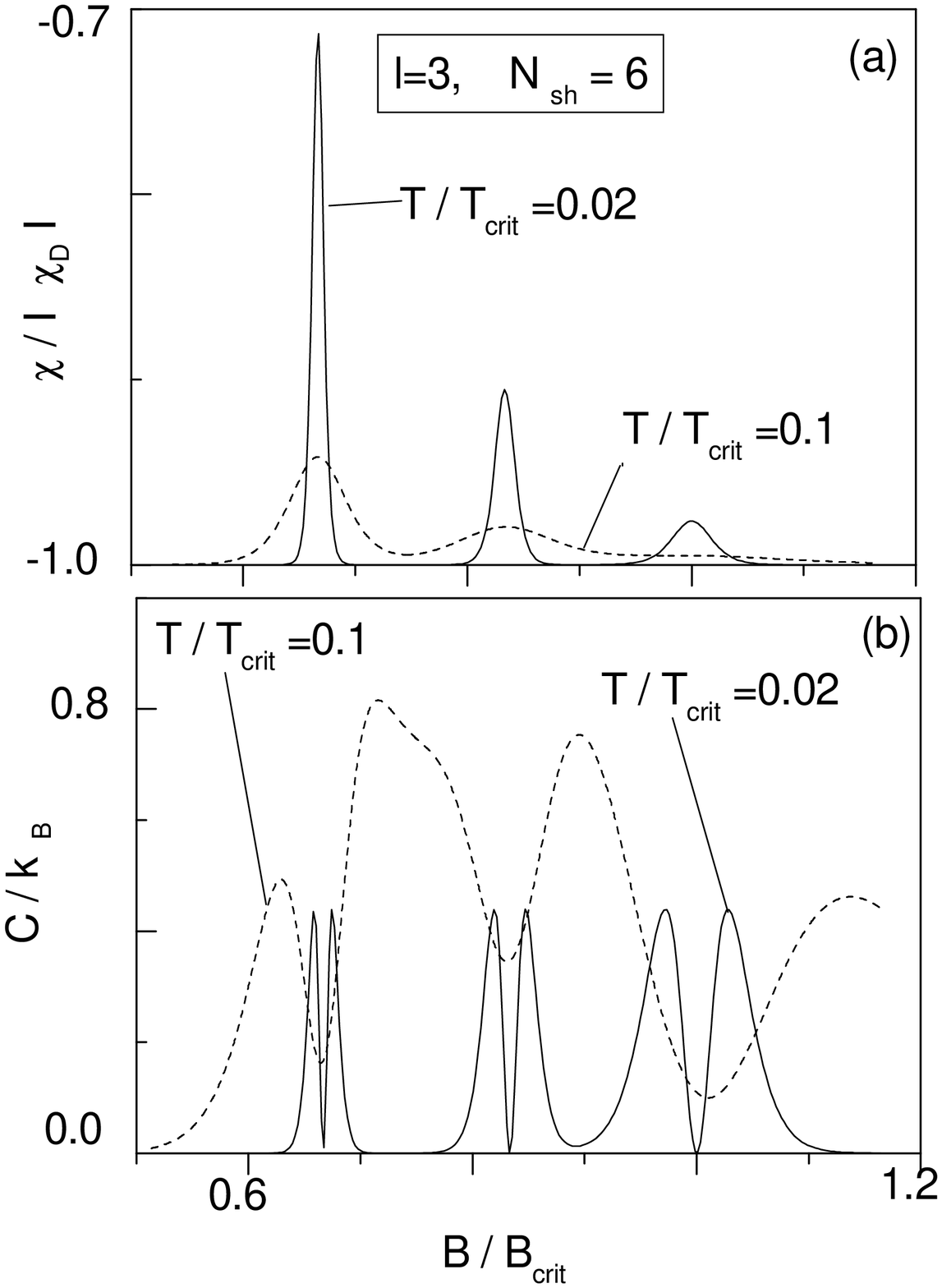}} \caption {Magnetic
oscillations in a superconducting sphere. $B_{\mathrm{crit}}$ is
the critical magnetic field of the BCS theory at $T=0$,
$\Delta_{\mathrm{BCS}}(T=0,B_{\mathrm{crit}})=0$. (a)~$\chi$ vs
$B$ ($\chi_{\mathrm{D}}$ is the diamagnetic susceptibility for an
$Al$--cluster with~$N=10^3$~). (b)~$C$ vs $B$ for the same
cluster.}
\end{figure}
Oscillations of $\chi$ and $C$ created
by level crossings in SSM spread on a range of rather small
fields $\omega\simeq\Delta_{\mathrm{BCS}}(0)/\widetilde l_F$
\cite{kuzmenko}
($B<0.5$ $T$ for $Al$-clusters with $N>10^3$ and $m=m^*$).
In nonspherical superconducting clusters where a weak magnetic
field does not influence orbital motion, the level crossings are
determined only by alignment of spins. Therefore greater values of
$B_{\mathrm{min}}$ are expected to induce the first extrema in $\chi$
and $C$.
\section{Conclusion}
In conclusion we have shown that an increasing uniform magnetic
field applied to a mesoscopic system transforms the structure of
the ground state in a such a way that in level crossing points
synchronously appear local extrema in magnetic susceptibility
({\em maxima\/}) and heat capacity ({\em minima\/}). Thereby these
thermodynamic quantities oscillate with growth field. Periodicity
and amplitudes of the low temperature oscillations carry
information concerning properties of the quantum states of
investigated normal and superconducting systems.

\end{document}